\documentclass[runningheads,a4paper]{llncs}

\usepackage{wrapfig}
\usepackage{amsmath}
\usepackage{amssymb}
\usepackage{graphicx}
\usepackage{url}
\usepackage{paralist,listings}
\usepackage{color}

\newenvironment{list1}{\begin{list}{$\bullet$}
{\topsep 0 pt \parsep 0 pt \partopsep 0 pt \itemsep 0
pt}}{\end{list}}

\newcommand{\comment}[1]{}

\usepackage{times}

\begin{document}

\title{\textsf{EFSMT}: A Logical Framework for Cyber-Physical Systems}

\author{
Chih-Hong Cheng\inst{1}
\and
Natarajan Shankar\inst{2}
\and
Harald Ruess\inst{1}
\and
Saddek Bensalem\inst{3}
}

\authorrunning{Chih-Hong Cheng, Natarajan Shankar,  Harald Ruess and Saddek Bensalem}

\institute{
fortiss GmbH, Guerickestr. 25, 80805 M{\"u}nchen, Germany\\
\and
SRI International,  333 Ravenswood,  Menlo Park, CA 94025, United States\\
\and
Verimag Laboratory, 2, avenue de Vignate, 38610 Gi\`{e}res, France
}
\maketitle


\begin{abstract}

The design of cyber-physical systems is challenging in that it
includes  the analysis and synthesis of distributed and embedded real-time systems
for controlling, often in a nonlinear way, the environment.
We address this challenge with \textsf{EFSMT},
the exists-forall quantified first-order fragment of propositional combinations over
constraints (including nonlinear arithmetic), as the logical framework and foundation
for analyzing and synthesizing cyber-physical systems.
We demonstrate the expressiveness of \textsf{EFSMT} by reducing a number of  pivotal  verification
and synthesis problems to \textsf{EFSMT}.
Exemplary problems in this paper include  synthesis for robust control via BIBO stability,
Lyapunov coefficient finding for nonlinear control systems,
distributed priority synthesis for orchestrating system components,
and synthesis for  hybrid control systems.
We are also proposing an algorithm for solving \textsf{EFSMT} problems based
on the interplay between two SMT solvers  for respectively solving universally and existentially
quantified problems. This algorithms builds on commonly used techniques in modern SMT solvers,
and generalizes them to  quantifier reasoning by counterexample-guided constraint strengthening.
The \textsf{EFSMT} solver uses Bernstein polynomials
for solving nonlinear arithmetic constraints.
\end{abstract}

\vspace{-2mm}

\section{Introduction}

The design of cyber-physical systems is challenging in that it includes the analysis and synthesis of distributed and embedded real-time systems for controlling nonlinear environments. We address this challenge by proposing \textsf{EFSMT}, a verification and synthesis engine  for solving  exists-forall quantified propositional combinations of constraints,  including nonlinear arithmetic.
Expressiveness and applicability of the \textsf{EFSMT} logic and solver is demonstrated by means of reducing a
number of  pivotal verification and  synthesis problems for cyber-physical systems to this fragment of first-order
arithmetic.

Over the last years, many verification tasks have been successfully reduced
to satisfiability problems in propositional logics  (SAT) extended with constraints
in rich  combinations of theories, and satisfiability modulo theory (SMT) solvers are
predominantly used for many software and system verification tasks.
Among many others, SMT has  been used  for optimal task
scheduling~\cite{yuan2008hardware,steiner2010evaluation},
bounded model checking for timed automata~\cite{Sorea:mtcs02} and infinite systems~\cite{dMRS02},
the detection of concurrent errors~\cite{lahiri2009static}, and
behavioral-level planning~\cite{ernst1997automatic,kautz1999unifying}\@.
The main attraction of these reductions lies in the fact that the original verification
and synthesis problems benefit from advances in research and technology for
solving SAT and SMT problems. In particular, it is very hard (and tedious) to
outperform search heuristics of modern SAT solvers or the combination of decision procedures in SMT solvers.
These logical reductions however are not a panacae and often need to be complemented with
additional structural analysis, since useful structural information is often \emph{lost in reduction}\@.

In this paper, we show that many different design problems for cyber-physical systems
naturally reduce to \textsf{EFSMT}, an exists-forall quantified fragment of first-order logic, which
includes nonlinear arithmetic.
Universally quantified variables are used for modeling uncertainties, and the
search for design parameters is equivalent to finding appropriate assignments for the
existentially bound  variables.
In this way  we show that  \textsf{EFSMT}  is expressive enough to encode a large variety
of design, analysis and synthesis tasks for cyber-physical systems including
\begin{list1}
\item Synthesis for robust control via BIBO stability;
\item Lyapunov coefficient finding for nonlinear control systems;
\item Distributed priority synthesis for orchestrating system components; and
\item Synthesis for hybrid control systems.
\end{list1}

We are proposing an optimized verification engine for solving \textsf{EFSMT} formulas,
which is based on the interplay of two \textsf{SMT}~\cite{cimatti2008beyond} solvers for formulas
of different polarity as determined by the top-level exists-forall quantifier alternation.
The basic framework for combining two propositional solver and exchanging
potential witnesses and counter-examples for directing the search.
We lift their basic procedure to the \textsf{EFSMT} logic and propose a number
of optimizations, including so-called extrapolation, which is inspired by the concept of widenings in abstract
interpretation. The \textsf{EFSMT} engine also incorporates a novel decision procedure~\cite{cheng:JBernstein:2013}
based on Bernstein polynomials for solving propositional combinations of non-linear arithmetic
constraints. Our developed arithmetic verification engine is promising in that it outperforms commonly used solvers based on
cylindrical algebraic decomposition by at least one or two orders of magnitude on our benchmark examples.
The implementation of  \textsf{EFSMT}  is based on Yices2~\cite{dutertre2006yices} and
JBernstein\cite{cheng:JBernstein:2013};
it is currently being integrated  into the Evidential Tool Bus~\cite{springerlink:10.1007/11576280_3,etb2013}\@.

The main contributions of this paper are
(1) the design and implementation of an optimized \textsf{EFSMT} solver based on established \textsf{SMT} solver technology, and
(2) presented reductions of a variety of design, analysis, and synthesis tasks for cyber-physical systems to logical problems in \textsf{EFSMT}\@.
Therefore the logical framework \textsf{EFSMT} represents a unified approach for diverse design problems,
and may be considered to be the logical equivalent of a swiss-army knife for designing cyber-physical systems.

The rest of the paper is structured as follows.  We describe the
exists-forall problem in Section~\ref{sec.prelim} and the underlying
algorithm of \textsf{EFSMT} in Section~\ref{sec.algorithm}.
Section~\ref{sec.non.linear} presents different methods used in \textsf{EFSMT} for solving problems in nonlinear
real arithmetic and apply them on some case studies.
Section~\ref{sec.examples} includes various reductions of design problems to \textsf{EFSMT} problems.
The implementation and programming interface of  \textsf{EFSMT} is outlined in~Section~\ref{sec.evaluation}\@.
We state related work in Section~\ref{sec.related} and  conclude with~Section~\ref{sec.conclusion}\@.

\vspace{-2mm}
\section{Preliminaries}\label{sec.prelim}

Let $\overline{x}$, $\overline{y}$ be a vector of $m$ and $n$ disjoint
variables. The general form of \emph{exists-forall problems} is
represented in Eq.~\ref{eq.forall.general}, where $[\overline{l_{x}},
\overline{u_{x}}] = [l_{x_1}, u_{x_1}]\times\ldots\times [l_{x_m},
u_{x_m}] \subseteq \mathbb{Q}^{m}$ and $[\overline{l_{y}},
\overline{u_{y}}] = [l_{y_1}, u_{y_1}]\times\ldots \times[l_{y_n},
u_{y_n}]\subseteq \mathbb{Q}^{n}$ is the \emph{domain} for
$\overline{x}$ and $\overline{y}$. $\phi(\overline{x}, \overline{y})$
is a quantifier-free formula that involves variables from
$\overline{x}$ and $\overline{y}$ of \textsf{boolean},
\textsf{integer}, fixed-point numbers (finite width), or \textsf{real}. We assume that the formula is well-formed,
i.e. it evaluates to either \textsf{true} or \textsf{false} provided
that all variables are assigned. Therefore, we do not require all variables to have the same type.

\vspace{-2mm}
\begin{equation}\label{eq.forall.general}
\exists \overline{x} \in [\overline{l_{x}}, \overline{u_{x}}] \; \forall \overline{y} \in [\overline{l_{y}}, \overline{u_{y}}]:
\phi(\overline{x}, \overline{y})
\end{equation}

$\phi(\overline{x}, \overline{y})$ is composed from a propositional combination of
\begin{inparaenum}[(a)]
\item boolean formula,
\item linear arithmetic for integer variables, and
\item linear and nonlinear polynomial constraints for real
  variables.
\end{inparaenum}
This combination enables the framework to model
discrete control in the computation unit (e.g., CPU),
physical constraints in the environment, and constraints of device
models. Integer-valued variables are used to encode locations and discrete control,
whereas the two-valued Boolean domain $\{0,1\}$ is used for encoding switching logic.
Boolean operations are encoded in arithmetic in the usual way, that is  $x_1\vee x_2$,
$x_1\wedge x_2$, and $\neg x_1$ are encoded, respectively, by $x_1+x_2$, $x_1 x_2$ and $1-x_1$.
This choice of interpretations is influenced by the requirements for the synthesis problems
considered in this paper. However, the solvers described below can easily be extended to
work with the rich combination of theories usually considered in SMT solving.
Notice also that constraints involving trigonometric functions are sometimes
encoded in terms of polynomials with an extra universal and real-valued variable $z$ for stating
conservative error estimates.

The following formula is an exist-forall problem.
\begin{multline} \label{ex:simple}
(\exists x \in [-30,30]\cap\mathbb{R})~(\forall y \in [-30, 30]\cap\mathbb{R}) (0<y<10) \rightarrow (y-2x<7)
\end{multline}

%

%

\vspace{-2mm}
\section{Solving \textsf{EFSMT}}\label{sec.algorithm}

\begin{wrapfigure}{r}{0.55\textwidth}
 \includegraphics[width=0.53\columnwidth]{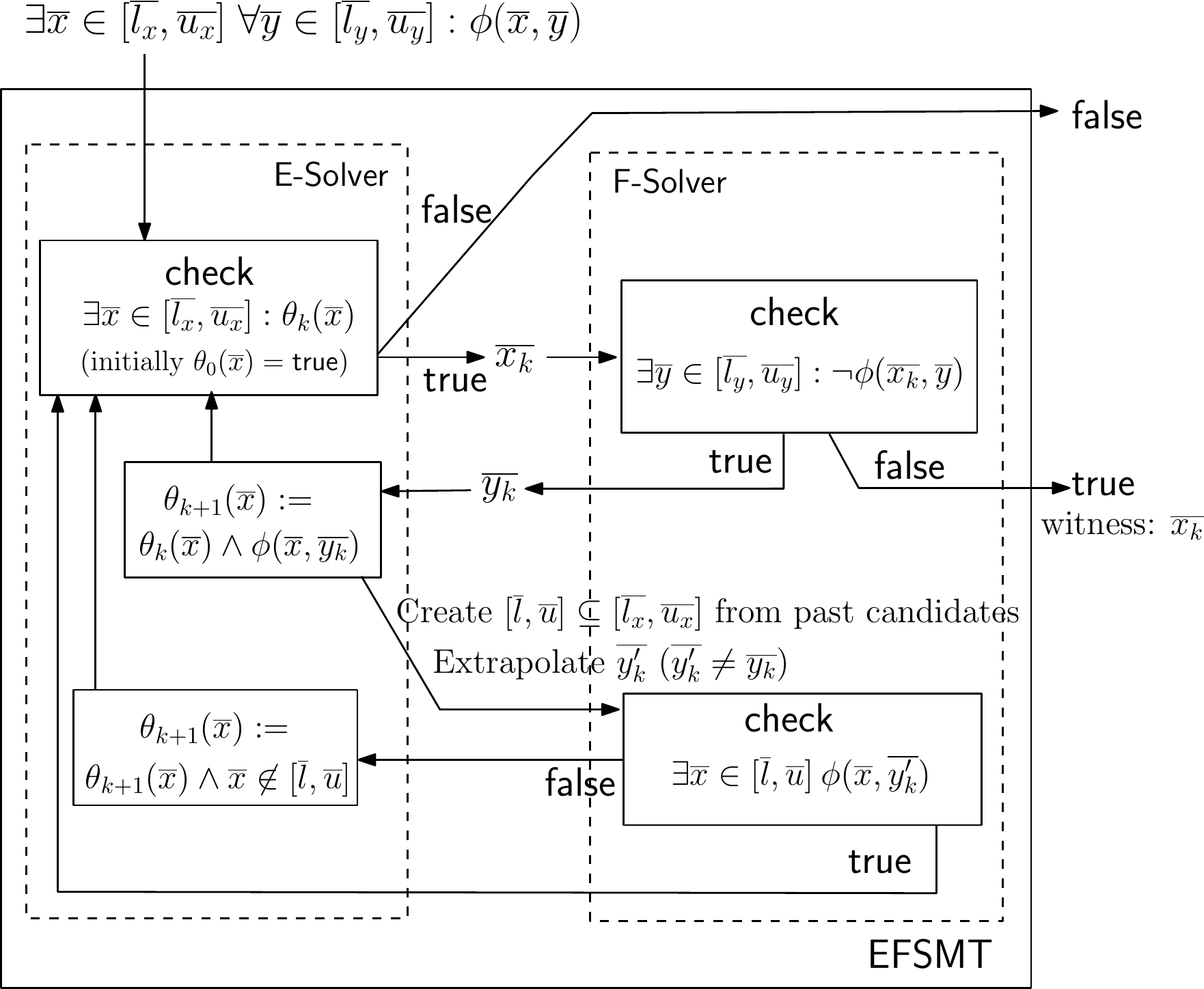}
  \caption{Algorithmic flow for \textsf{EFSMT}.}
 \label{fig:algorithm}
\end{wrapfigure}

We outline a verification procedure for solving  \textsf{EFSMT} problems of the form
 $$(\exists \overline{x} \in [\overline{l_{x}}, \overline{u_{x}}])\;
                (\forall \overline{y} \in [\overline{l_{y}}, \overline{u_{y}}])\;\phi(\overline{x}, \overline{y})\mbox{.}$$
This procedure relies on SMT solvers for deciding the satisfiability of propositional combinations of constraints (in a given theory)\@.
If the input formula is unsatisfiable the SMT solver returns \textsf{false}; otherwise it is assumed to return \textsf{true} together with
a satisfying variable assignment.
The solver in Figure~\ref{fig:algorithm} is based on two instances, the so-called \textsf{E-solver} and  \textsf{F-solver} of such SMT solvers.
These two solvers are applied to quantifier-free formulas of different polarities in order to reflect the quantifier alternation, and they are
combined by  means of a counter-example  guided refinement strategy.

\paragraph{Counterexample-directed search.}
A straightforward method for solving \textsf{EFSMT} is to guess a variable assignment,
say $\overline{x_{0}}$, and to verify  that the sentence
$(\forall \overline{y})\phi(\overline{x_{0}}, \overline{y})$ holds\@.
The \textsf{F-solver} may be used to decide validity problems of the form
 $(\forall \overline{y})\psi(\overline{y})$ by reducing them to the unsatisfiability problem
for $(\exists \overline{y})\neg\psi(\overline{y})$\@.

In this way, for Eq.~\ref{ex:simple}, after guessing the assignment $x := 0$ for the \textsf{EFSMT} constraint, the problem
is reduced to the validity problem for  $\forall y \in [-30, 30]: (0<y<10) \rightarrow (y<7)$ - which obviously fails to hold\@.
Instead of blindly guessing new instantiations, one might use failed proof attempts and counter-examples $y_0$ provided by
the \textsf{F-solver} to restrict the search space for assignments to the existential variables and to guide the selection of new assignments.
If  the  \textsf{F-solver} generates, say,  the counter example $y := 9$, then $((0<9<10) \rightarrow (9-2x<7))$, which is equivalent to
$x>1$, is passed to the \textsf{E-solver}\@. Using this constraint, the \textsf{E-solver} has cut its search space in half\@.


The counterexample-guided verification procedure for \textsf{EFSMT} based on two SMT solvers  \textsf{E-solver} and \textsf{F-solver}
is illustrated in the upper part of Fig.~\ref{fig:algorithm}\@.
At the $k$-th iteration, the \textsf{E-solver} either generates an instance $\overline{x_k}$ for $\overline{x}$
or the procedure returns with \textsf{false}\@.
An $\overline{x_k}$ provided by the \textsf{E-solver} is passed to the \textsf{F-solver}
for checking if $\exists \overline{y} \in [\overline{l_{y}}, \overline{u_{y}}]:\neg\phi(\overline{x_k}, \overline{y})$ holds.
In case there is a satisfying assignment $\overline{y_k}$,  the \textsf{F-solver} passes the constraint $\phi(\overline{x}, \overline{y_k})$ to the \textsf{E-solver},
for ruling out such $\overline{x}$ as potential witnesses.
Future cancidate witnesses  $\overline{x_{k+1}}$ should therefore not only   $\phi(\overline{x_{k+1}}, \overline{y_0}), \ldots, \phi(\overline{x_{k+1}}, \overline{y_{k-1}})$ but also
 $\phi(\overline{x_{k+1}}, \overline{y_k})$ returns \textsf{true}\@.



\paragraph{Logical contexts.}
SMT solvers such as Yices or Z3~\cite{de2008z3} support \emph{logical contexts}, that is,
finite sequences of  conjoined contextual constraints,  together with operations for
dynamically \textsf{push}ing and \textsf{pop}ping constraints as the basis for
efficiently implementing backtracking search.
The \textsf{EFSMT} procedure uses  these contextual operations in order to
avoid the re-processing of formulas by the \textsf{F-solver}\@.
Considering again our running example, the  \textsf{F-solver} pushes the following contextual
information: $(0<y<10) \wedge (y-2x\geq 7)$\@.
Whenever an assignment $x := x_i$ is generated by the \textsf{E-solver},
a new constraint $x=x_i$ is pushed  and  satisfiability of the constraint
     $(x=x_i) \wedge ((0<y<10) \rightarrow  (y-2x\geq7))$ is being checked.
Then, the solver pops the context to recover
      $(0<y<10) \rightarrow (y-2x\geq7)$
and awaits the next candidate assignment
     $x=x_{i+1}$\@.
Likewise, the \textsf{E-solver} pushes the  constraints generated by the \textsf{F-solver}\@.

\paragraph{Partial Assignments.}
Some SMT solvers such as Yices and Z3 provide partial variable assignments.
If a variable $x$ is not in the codomain of such a partial assignment, then every possible interpretation of $x$ yields a
satisfying assignment\@.
In this way, the \textsf{EFSMT} procedure utilizes partial variable assignments of the \textsf{F-solver}
for speeding up convergence by further decreasing the search space for candidate witnesses for the \textsf{E-solver} in
every iteration.  Symbolic counterexamples, such as $7 \le y < 10$ in our
running example, have the potential of accelerating convergence even more.


\paragraph{Extrapolation.}
Given a subspace $[\overline{l}, \overline{u}] \subseteq [\overline{l_{x}}, \overline{u_{x}}]$ and
 $\overline{y_k'} \in [\overline{l_{y}}, \overline{u_{y}}]$\@.
 If the formula $\forall  \overline{x}\in [\overline{l}, \overline{u}]: \neg\phi(\overline{x}, \overline{y_k'})$ holds
then any $x \in [\overline{l}, \overline{u}]$ can be ruled out as a candidate witness.
This subspace elimination process is described in the bottom part of Fig.~\ref{fig:algorithm},
where the \textsf{F-solver} checks the negated property
   $(\exists  \overline{x}\in [\overline{l}, \overline{u}])\,  \phi(\overline{x}, \overline{y_k'})$\@.
The infeasibility test appears when the solver continuously tries to refine a relatively small subspace
without finding a satisfactory solution.
Notice that extrapolation technique is similar to \emph{widening} in abstract interpretation~\cite{cousut77}\@.

The generation of $\overline{y_k'}$ is based on extrapolation, as
shown in the following example: $(\exists x\in[0,10])(\forall y\in
[0,10]) y\geq x$.
The formula evaluates to \textsf{true} with witness $x=0$\@.
Without extrapolation the \textsf{E-solver} produces the sequence
 $2, \frac{1}{2}, \frac{1}{8}, \frac{1}{32}\ldots$ of candidate witnesses,
and the \textsf{F-solver} produces the sequence $1,
\frac{1}{4}, \frac{1}{16}, \frac{1}{64}\ldots$ of counterexamples.
To achieve termination, the solver observes the convergence of $x$ and generates
$(0,\frac{1}{32}]$ and extrapolates $y$ to be $0$, therefore $(\exists x\in(0,\frac{1}{32}]) 0>x$ is \textsf{false}.
Therefore, after checking the constraint generated by extrapolation, the \textsf{E-solver} rules out
all values  greater than 0, and the remaining value $0$ is the witness.

\paragraph{Incompleteness for existential reals; completeness for fixed-point numbers.}
The \textsf{EFSMT} procedure in Figure~\ref{fig:algorithm} is sound in that it returns \textsf{true}
only in case the  input sentence holds and \textsf{false} only in cases it does not hold.
Not too surprisingly, the \textsf{EFSMT} procedure as stated above, is incomplete, as demonstrated by a simple example:
      $$(\exists x\in[0,10]\cap \mathbb{R})\; (\forall y\in [0,10]\cap \mathbb{R}): x>0 \wedge ((y>0 \wedge y\neq x) \rightarrow y>x)$$
\textsf{EFSMT} should return  \textsf{false}\@.
However, the \textsf{E-solver} produces the sequence  $2, \frac{1}{2}, \frac{1}{8}, \frac{1}{32}\ldots$ of guesses,
whereas the \textsf{F-solver} produces counter-examples $1, \frac{1}{4}, \frac{1}{16},
\frac{1}{64}\ldots$\@.
Every counter-example $y_k$ shrinks the search space by posing an additional constraint $x<y_k$ to the
\textsf{E-solver}, but the added restriction is not sufficient for the procedure to conclude \textsf{false}\@.
In this case, extrapolation as described above is not helpful either.

However, the incompleteness only comes with existential variables having domain over reals or rationals. As existential variables
are used as design parameters that needs to be synthesized, in many cases we pose additional requirements to have existential variables 
be representable as integers or fixed-point numbers. In this way, the method at the worst case only searches for all possible scenarios, and completeness is guaranteed.

%

%

\vspace{-2mm}
\section{Handling Nonlinear Real Arithmetic}\label{sec.non.linear}

One of the main challenges for the \textsf{EFSMT} verification procedure
 is the design of  an efficient and  reliable \emph{little engine}  for solving nonlinear constraints.
We are describing three such solving techniques in \textsf{EFSMT} which prove to be particularly useful.

\paragraph{Linearization.}
Many nonlinear arithmetic constraints naturally reduce to linear constraints in the \textsf{EFSMT} algorithm in Figure~\ref{fig:algorithm}\@.
Consider, for example, the constraint $(\exists s, t)(\forall y, z) sy+2t+tz>0$\@.
Using the assignment $s:=s_0, t:=t_0$, with $s_0$, $t_0$ constants,
the \textsf{F-solver} determines the formula $(\forall y, z) s_0y+2t_0+t_0z>0$ in
linear arithmetic.
Now, assume that the  \textsf{F-solver} returns $y:=y_0, z:=z_0$ as a
 witness for $(\exists y, z) s_0y+2t_0+t_0z\leq 0$\@.
Then the linear constraint $sy_0+2t+tz_0 = (y_0)s+(z_0+2)t>0$ is supplied to the \textsf{E-solver}.
In particular, constraints are linearized when every monomial has at most two variables, one of which is existentially and the other universally bound\@.

\paragraph{Bitvector Arithmetic.}
The second approach to nonlinear arithmetic involves constraint strengthening techniques and subsequently, finding a witness for the strengthened constraint with \emph{bitvector arithmetic}.
Bitvector arithmetic presents every value with only finitely many bits, similar to the \emph{fixed-point representation}, and is therefore only approximate.
A bit-vector representation supports nonlinear arithmetic by allowing arbitrary multiplication of variables.
Solving exists-forall constraints with bitvector arithmetic is implemented in \textsf{EFSMT} as an extension based on 2QBF\@.

Let $[\overline{l}, \overline{u}]_{BV}$ be the set of points in $[\overline{l}, \overline{u}]$ that can be represented by bitvectors. Intuitively, for constraint $\exists \overline{x} \in [\overline{l_{x}}, \overline{u_{x}}] \; \forall \overline{y} \in [\overline{l_{y}}, \overline{u_{y}}]: \phi(\overline{x}, \overline{y})$, a positive witness $\overline{x_k}_{BV}$ for bitvector arithmetic constraint $\exists \overline{x} \in [\overline{l_{x}}, \overline{u_{x}}]_{BV} \; \forall \overline{y} \in [\overline{l_{y}}, \overline{u_{y}}]_{BV}: \phi(\overline{x}, \overline{y})$ is not necessarily a solution for the original problem, as $\overline{x_k}_{BV}$ does not consider points within $[\overline{l_{y}}, \overline{u_{y}}] \setminus [\overline{l_{y}}, \overline{u_{y}}]_{BV}$. However, $\overline{x_k}_{BV}$ can be a solution for the original problem when there exists a proof stating that checking bitvector points $[\overline{l_{y}}, \overline{u_{y}}]_{BV}$ is equivalent to checking the whole interval $[\overline{l_{y}}, \overline{u_{y}}]$.
To achieve this goal, one can \emph{strengthen} the original quantifier-free formula $\phi(\overline{x}, \overline{y})$ to another formula $\phi'(\overline{x}, \overline{y})$.


We use the following example $\exists k_p k_i \,\forall t, u\in [0, 10]: k_pk_i>tu$ to explain the strengthening approach.
In bitvector arithmetic, let $1_{bit}$ be the smallest unit for addition. Given a bitvector variable with value $t_0$, its successor bitvector value is $t_0+1_{bit}$, and any value $s\in \mathbb{R}$ in between is $t_0+\kappa_{t} 1_{bit}$, where $0< \kappa_t<1$.
Therefore, when using bitvector arithmetic in \textsf{EFSMT} on the following strengthened problem:
\begin{equation}
\exists k_{p} k_{i}\, \forall t, u \in [0, 10]_{BV}: k_{p}k_{i}>(t+1_{bit})(u+1_{bit})
\end{equation}
a witness $(k_{p}, k_{i}) = (a, b)$ is also a witness for $\exists k_p k_i \,\forall t, u\in [0, 10]: k_pk_i>tu$ where variables $k_p, k_i, t, u$ range over reals. This is because $(t+1_{bit})(u+1_{bit}) > (t+\kappa_{t} 1_{bit})(u+\kappa_{u} 1_{bit})$, for $0< \kappa_t, \kappa_u <1$.

Strengthening is a powerful technique, but finding an ``appropriate''
strengthened condition may require human intelligence. Consider for
example, $\exists \overline{x} \in [\overline{l_{x}},
\overline{u_{x}}]_{BV} \; \forall \overline{y} \in [\overline{l_{y}},
\overline{u_{y}}]_{BV}:\textsf{false}$, which is equivalent to
\textsf{false}. This is a strengthened constraint, as $(\textsf{false}
\rightarrow \exists \overline{x} \in [\overline{l_{x}},
\overline{u_{x}}] \; \forall \overline{y} \in [\overline{l_{y}},
\overline{u_{y}}]: \phi(\overline{x}, \overline{y})) \equiv
\textsf{true}$, but is of no interest since the strengthened condition
can not be proved \textsf{true}.

\paragraph{Bernstein polynomials.}
The nonlinear solving techniques described so far rely on features of current SMT solvers (e.g., linear arithmetic, bitvectors)\@.
In contrast, we are now describing a customized \textsf{F-solver} for nonlinear real arithmetic based on Bernsteinst polynomials.
This requires to restrict ourselves to propositional constraints of \emph{assume-guarantee} form.
\begin{multline}\label{eq.forall.assume.guarantee}
(\exists \overline{x} \in [\overline{l_{x}}, \overline{u_{x}}]) \; (\forall \overline{y} \in [\overline{l_{y}}, \overline{u_{y}}]) 
\bigwedge^k_{j=1}  ((\bigwedge^{q}_{p=1} \rho_{jp}(\overline{x}, \overline{y})\,\textsf{op}_{jp}\,d_{jp})
\rightarrow \varphi_j(\overline{x}, \overline{y})\,\textsf{op}_j\,e_j)
\end{multline}
where
(1) $\rho_{jp}, \varphi_j$ are polynomials over real variables in $\overline{x}, \overline{y}$ and
(2) $\textsf{op}_{jp}, \textsf{op}_{j}\in \{>,\geq,<,\leq\}$\@.
\textsf{JBernstein} is a polynomial constraint checker based on Bernstein polynomials~\cite{MN12} that checks properties of the form $\forall \overline{y} \in [\overline{l_{y}}, \overline{u_{y}}]: \bigwedge^k_{j=1}  ((\bigwedge^{q}_{p=1} \rho_{jp}(\overline{y})\,\textsf{op}_{jp}\,d_{jp})
\rightarrow \varphi_j(\overline{y})\,\textsf{op}_j\,e_j)$, i.e., Eq.~\ref{eq.forall.assume.guarantee} without existential variables.
Here, \textsf{JBernstein} is used as an \textsf{F-solver}\@.

The algorithm of the Bernstein approach consists of three steps:
\begin{inparaenum}[(a)]
\item range-preserving transformation,
\item transformation from polynomial to Bernstein basis, and
\item a sequence of subspace refinement attempts until a proof is
  found or the number of refinement attempts exceeds a threshold.
\end{inparaenum} As a quick illustration, consider $\forall x\in [1,3]: \phi(x)= x^2-4x+4>-3$. The range-preserving transformation performs linear scaling so that every variable after translation is in domain $[0,1]$ but the range remains the same; in this example by setting $y=\frac{x-1}{2}$ we derive $\forall y\in [0,1]: \phi'(y)= 4y^2-4y+1> -3$. $\phi'(y)$ has polynomial basis $\{y^2, y, 1\}$. $\phi'(y)$ can also be rewritten as $\textbf{1}({}_2^0)(1-y)^2 \textbf{-2}({}_2^1)y(1-y)+\textbf{1}({}_2^2)(y)^2$, where $\{({}_2^k)y^k(1-y)^{2-k}|k=0,1,2\}$ is the Bernstein basis. To check if $4y^2-4y+1>-3$ holds for all $y\in[0,1]$, it is sufficient to show that all coefficients in the Bernstein basis are greater than $-3$. Since $\textbf{1}>-3$ and $\textbf{-2}>-3$, the property holds.

For {\small$\forall \overline{y} \in [\overline{l_{y}}, \overline{u_{y}}]: \bigwedge^k_{j=1}  ((\bigwedge^{q}_{p=1} \rho_{jp}(\overline{y})\,\textsf{op}_{jp}\,d_{jp})
\rightarrow \varphi_j(\overline{y})\,\textsf{op}_j\,e_j)$}, \textsf{JBernstein} checks the condition by examining if every assume-guarantee rule $(\bigwedge^{q}_{p=1} \rho_{jp}(\overline{y})\,\textsf{op}_{jp}\,d_{jp})\rightarrow \varphi_j(\overline{y})\,\textsf{op}_j\,e_j$ holds. Every assume-guarantee rule $\alpha \rightarrow \beta$ is discharged into its disjunction form $\alpha\vee \beta$. $\alpha\vee \beta$ holds if every subspace satisfies either $\alpha$ or $\beta$, and $\alpha\vee \beta$ fails if exists a point in the subspace that violates $\alpha$ and $\beta$.

The Bernstein polynomial checker supports linearization as follows.
Consider, for example,  the  constraint $\exists x, z \in [-10,10]: \forall y\in [-10,10]: xy^2+4zy+x+5>0$\@.
Here,  the \textsf{E-solver} may only use linear arithmetic whereas the \textsf{F-solver} uses \textsf{JBernstein}.
Moreover, one may also restrict the search of the \textsf{E-solver}  for witnesses to those which may be encoded using bitvectors. 
\vspace{-2mm}
\section{Reductions to \textsf{EFSMT}}\label{sec.examples}

We illustrate the expressive power of \textsf{EFSMT} logical framework
by reducing a variety of design problems for cyber-physical systems to this
fragment of logic.

\subsection{Safety Orchestration for Component-based Systems}

\begin{wrapfigure}{r}{0.4\textwidth}
 \includegraphics[width=0.35\columnwidth]{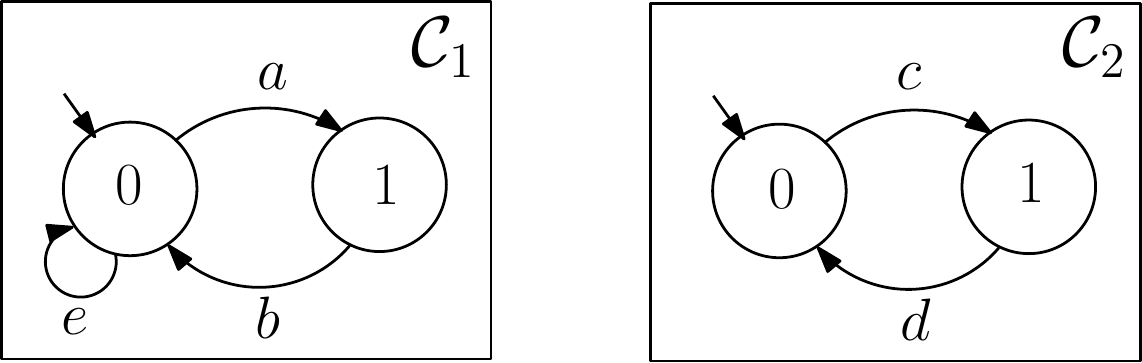}
  \caption{A simple component system.}
 \label{fig:priority.synthesis}
\end{wrapfigure}

We first present an encoding technique that synthesizes glue code for safety orchestration problems in component-based systems.

\paragraph{Problem Description.} Consider the sample system in
Fig.~\ref{fig:priority.synthesis} that includes two components
$\mathcal{C}_1$ and $\mathcal{C}_2$. Each edge corresponds to an
action. For actions $a$ and $c$, the components move from state $0$ to
state $1$ and start consuming a resource. Actions $b$ and $d$ release
the resource. In the initial state the two components do not consume
the resource. Since the resource usage is exclusive, the state $(1,1)$
is considered a risk state.

Clearly, it is possible to reach state $(1,1)$ from the initial state
$(0,0)$. Therefore, suitable orchestration is needed. However, the
orchestration should guarantee global progress and never introduce new
deadlocks. For example, blocking any execution from the initial state
eliminates the possibility to reach $(1,1)$ but is undesirable since
none of the components can use the resource.

The orchestration mechanism is restricted to a set
$S=\{\alpha\prec\beta\, | \,\alpha, \beta \in \{a,b,c,d,e\}\}$ of
\emph{priorities}~\cite{basu2006modeling}. Intuitively, $\alpha\prec
\beta$ means that whenever both $\alpha$ and $\beta$ actions are
enabled, the orchestration prefers action $\beta$ over
$\alpha$. Elements within the introduced set should ensure
transitivity (i.e., $\alpha\prec \beta , \beta \prec \gamma \in S
\rightarrow \alpha\prec\gamma \in S$) and irreflexivity (i.e.,
$\alpha\prec\alpha \not\in S$) to generate unambiguous semantics for
system execution. Overall, the problem of \emph{priority synthesis} is
to define a set of priorities which guarantees (by priorities) that
the system under control is free of risk and deadlock.

\paragraph{Encoding.} To encode a priority synthesis problem into an exists-forall problem, our method is to introduce \emph{templates} where the union of valid templates forms a \emph{safety-invariant} of the system. A safety-invariant is a set of states that has the following properties:

\begin{enumerate}
    \item The initial state is within the safety-invariant.
    \item Risk state are excluded from the safety-invariant.
    \item For every state $s$ that is within the safety-invariant, if
      action $s\xrightarrow{\alpha}s'$ is legal (i.e., it is not
      blocked by another action $\beta$ due to priorities), then $s'$
      is contained in the safety-invariant.
\end{enumerate}

As each component only has two states $\{0,1\}$, for component $C_i$, we use one boolean variable $x_i$ to indicate its current state.
Here we use two templates $(\textsf{m}_{val}, \textsf{m}_1)$ and  $(\textsf{n}_{val}, \textsf{n}_1, \textsf{n}_2)$. The first template has two Boolean variables $\textsf{m}_{val}, \textsf{m}_1$. When $\textsf{m}_{val}$ is set to \textsf{true}, the first template is used. When $\textsf{m}_1$ is assigned to \textsf{true}, the set of states that is covered in this template is $(1, \textsf{-})$, where symbol "$\textsf{-}$" means \emph{don't-cares} and includes all possible states in $\mathcal{C}_2$.
For each template, we need to declare both the primed and the unprimed version. In summary, we declare the following Boolean variables when translating the problem into \textsf{EFSMT}.

As each component only has two states $\{0,1\}$, for each component
$C_i$, we use a Boolean variable $x_i$ to indicate the current state.
Here we use two templates $(\textsf{m}_{val}, \textsf{m}_1)$ and
$(\textsf{n}_{val}, \textsf{n}_1, \textsf{n}_2)$. The first template
has two Boolean variables $\textsf{m}_{val}, \textsf{m}_1$. When
$\textsf{m}_{val}$ is \textsf{true}, the first
template is used (i.e., the first variable is a guard enabling a template).
When $\textsf{m}_1$ is \textsf{true}, the set of states covered in this
template is $(1, \textsf{-})$, where symbol ``-'' denotes
\emph{don't-cares} and includes all possible states of
$\mathcal{C}_2$.  For each template, we need to declare both the
primed and the unprimed versions. In summary, we declare the following Boolean variables when
translating the problem in a suitable form for \textsf{EFSMT}.

\begin{enumerate}
\item For every priority $\alpha\prec\beta$, declare an existential
  variable $\underline{\alpha\prec\beta}$. When
  $\underline{\alpha\prec\beta}$ evaluates to \textsf{true}, we
  introduce priority $\alpha\prec\beta$ to restrict the behavior. In
  this example, 25 variables are introduced.

    \item Every state variable in the template together with its primed version are declared as existential variables. In this example, we use two templates and have in total 8 variables $\textsf{m}_{val}, \textsf{m}_1, \textsf{m}_1', \textsf{n}_{val}, \textsf{n}_1, \textsf{n}_2, \textsf{n}_1', \textsf{n}_2'$.
    \item Every state variable and its primed version are declared as universal variables. In this example, we need four variables $x_1, x_1', x_2, x_2'$.
\end{enumerate}
Altogether we obtain the following constraints.
\begin{itemize}
    \item The primed version and unprimed version of the invariant should be the same. In this example, we add clauses such as $(\textsf{m}_{1}\Leftrightarrow \textsf{m}_{1}'), (\textsf{n}_{1}\Leftrightarrow \textsf{n}_{1}'), (\textsf{n}_{2}\Leftrightarrow \textsf{n}_{2}')$.
    \item At least one template should be enabled. In this example, we add the clause $(\textsf{m}_{val} \vee \textsf{n}_{val})$.
    \item If a state is an initial state, it is included in some valid invariant. In this example, the initial state has an encoding $(\neg x_1 \wedge \neg x_2)$. We introduce the constraint
    \begin{multline*}
        ((\textsf{m}_{val} \wedge (\neg x_1 \wedge \neg x_2)) \rightarrow (x_1 \Leftrightarrow \textsf{m}_1)) 
        \,\vee\, 
        ((\textsf{n}_{val} \wedge (\neg x_1 \wedge \neg x_2)))
        \rightarrow ((x_1 \Leftrightarrow \textsf{n}_1) \wedge (x_2 \Leftrightarrow \textsf{n}_2)))
    \end{multline*}
    \item If a state is a risk state, then it is not included in any valid invariant. In this example, the risk state has an encoding $(x_1 \wedge x_2)$. We introduce the constraint
           \begin{multline*}
         (\textsf{m}_{val} \rightarrow \neg ((x_1 \wedge x_2) \rightarrow (x_1 \Leftrightarrow \textsf{m}_1))) 
        \,\wedge \,
        ((\textsf{n}_{val} \rightarrow \neg ((x_1 \wedge x_2)
        \rightarrow ((x_1 \Leftrightarrow \textsf{n}_1)\wedge (x_2 \Leftrightarrow \textsf{n}_2)))
    \end{multline*}
    E.g., for the first line, if $\textsf{m}_{val}=\textsf{true}$ and $x_1 \wedge x_2$ is \textsf{true}, the template should not be $\textsf{m}_1=\textsf{true}$, as this makes the $\neg ((x_1 \wedge x_2) \rightarrow (x_1 \Leftrightarrow \textsf{m}_1))$ \textsf{false}.
    \item Encode the transition by considering the effect of priorities. For example, the encoding of transition $a$ is $\textsf{tran}_a := \neg x_1 \wedge x_1' \wedge (x_2 \Leftrightarrow x_2')$, and the encoding of transition $c$ is $\textsf{tran}_c := (x_1 \Leftrightarrow x_1')\wedge  \neg x_2 \wedge x_2'$. The condition for $a$ and $c$ to hold simultaneously is $\textsf{cond}_{a,c}:= \neg x_1 \wedge \neg x_2$. Therefore, the condition that considers the introduction of priority $a\prec c$ is
        $(\underline{a\prec c} \wedge \textsf{cond}_{a,c}) \rightarrow \neg \textsf{tran}_a$.
    The above constraint states that if priority $a\prec c$ is used, then whenever $a$ and $c$ can be selected, we prefer $c$ over $a$ (by disabling $a$). Following this approach, we construct the transition system $\textsf{tran}_{prio}$ that takes the usage of priorities into account. $\textsf{in}_{x}$ is defined as
       \begin{multline*}
        (\textsf{m}_{val} \wedge (x_1 \Leftrightarrow \textsf{m}_1))
        \vee
        (\textsf{n}_{val} \wedge ((x_1 \Leftrightarrow \textsf{n}_1) \wedge (x_2 \Leftrightarrow \textsf{n}_2)))
          \end{multline*}
    That is, $\textsf{in}_{x}$ specifies the constraint where a state is within template $m$ or $n$. We also create $\textsf{in}_{x}'$ that uses variables in their primed version.
    Finally, introduce the following constraint to \textsf{EFSMT}:
    $(\textsf{in}_{x} \wedge \textsf{tran}_{prio}) \rightarrow \textsf{in}_{x}'$, which ensures the third condition of a legal safety-invariant.

    \item Introduce constraints on properties of the introduced priorities such as transitivity and irreflexivity. For example, introduce $\neg (\underline{a\prec a})$, $\neg (\underline{b\prec b})$, $\neg (\underline{c\prec c})$, $\neg (\underline{d\prec d})$, and $\neg (\underline{e\prec e})$ to ensure irreflexivity. The transitivity and irreflexivity for priorities enforce a partial order over actions.

\end{itemize}

The above encoding not only ensures that the system can avoid entering any risk states, a feasible solution returned by \textsf{EFSMT} also never introduces new deadlocks\footnote{In the analysis, we set all deadlock states that appear in the original system to be risk states.}. This is because a priority $\alpha \prec \beta$ only blocks $\alpha$ when $\beta$ is enabled, and precedences of actions forms a partial order. Therefore, the restriction of using priorities as orchestration avoids bringing another quantifier alternation to ensure global progress\footnote{In general, to ensure progress, one should use three layers of quantifier alternation by stating  (informally) that there \emph{exists} a strategy such that \emph{for every} safe state, there \emph{exists} one safe state that is connected by the synthesized strategy.}.

For this example, \textsf{EFSMT} returns \textsf{true} with $\textsf{m}_1=\textsf{false}$, $\textsf{n}_1=\textsf{true}$, and $\textsf{n}_2=\textsf{false}$, meaning that the safety-invariant constructed by two templates is $\{(0,0),(0,1),(1,0)\}$. The set of introduced priorities for system safety is $\{a \prec d, c \prec b\}$.

\begin{wrapfigure}{r}{0.5\textwidth}
 \includegraphics[width=0.45\columnwidth]{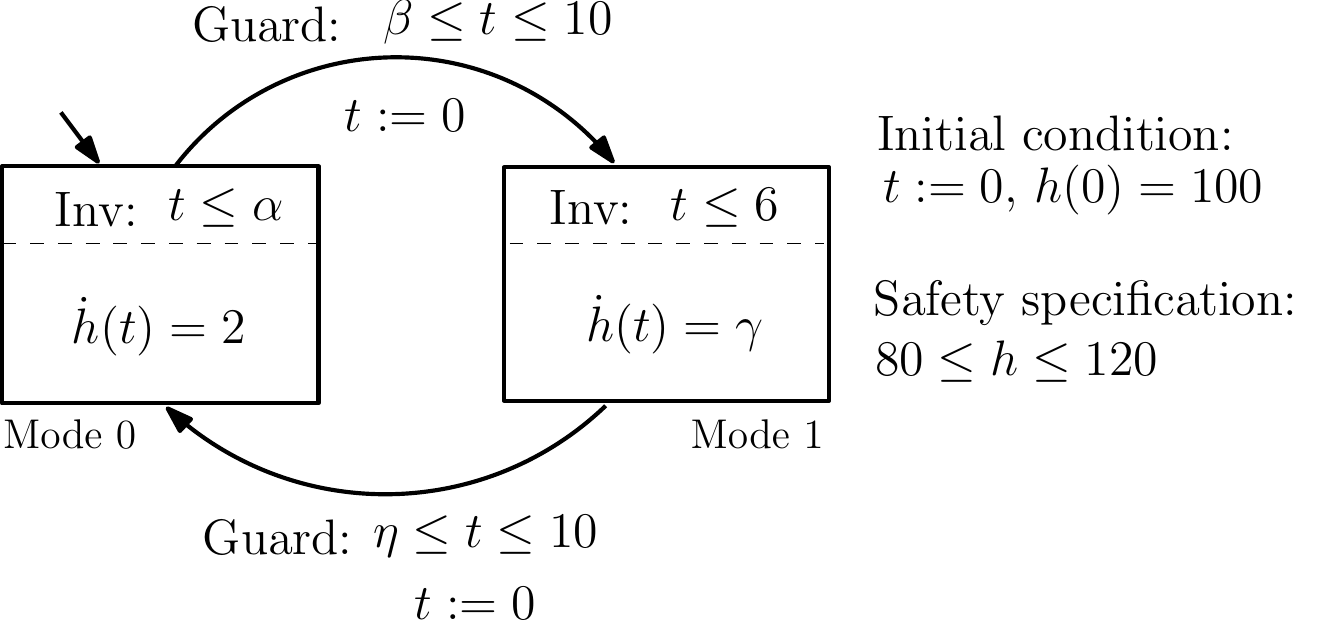}
 \vspace{0.5mm}
  \caption{A simple temperature control system.}
 \label{fig:temperature.control}
\end{wrapfigure}

\paragraph{Extensions.} When
components are considered independent execution units, priority
enforcement requires a communication channel. Consider for example the
priority $a\prec c$. Component $\mathcal{C}_1$ needs to observe
whether $C_2$ can execute $c$ in order to execute $a$ and conform to
the priority. Assume a unidirectional communication channel from $C_1$
to $C_2$. Such condition restricts the use of $a\prec c$ and
similarly, every usage of $\alpha\prec \beta$ where $\alpha \in
\{a,b,c\}$ and $\beta \in \{c, d\}$. When translating this requirement
into \textsf{EFSMT}, the solver only needs to introduce new
constraints $\neg (\underline{\alpha\prec \beta})$ to disable these
priorities. When the additional constraints, \textsf{EFSMT} returns
with $\textsf{m}_{val}=\textsf{true}, \textsf{m}_1=\textsf{false},
\textsf{n}_{val}=\textsf{false}$, meaning that only template
\textsf{m} is used with safe states $\{(0,0), (0,1)\}$.
For this example, priority $a\prec e$ is synthesized.

The encoding above can also be generalized to include \emph{knowledge} of each local component concerning
their respective view of global states. By introducing new existential variables, the solver can dynamically decide to use or ignore statically computed knowledge to guarantee safety. The encoding process is essentially the same, where  we additionally introduce constraints state that the use of knowledge can overcome the restriction due to communication.

\subsection{Timed and Hybrid Control Systems} \label{sub.sec.timed.systems}

The template-based techniques presented in the previous section can be
extended to the analysis of real-time control systems.
For simplicity we assume that timed systems only us one clock $t$\@.
A state is a pair $(s, t)$ where $s$ is the location and $t$ is the reading of the clock.
The safety-invariant ensures the following:

\begin{description}
\item[Initial state.] The initial state $(s_0, 0)$, where $s_0$ is the
  initial location, is within the safety-invariant.
\item[Risk states.] No risk state is within the safety-invariant.
\item[Progress of time.] For every state $(s, t)$ that is within the
  safety-invariant, if a $\delta$-interval time-progress $(s,
  t)\xrightarrow{\delta}(s, t+\delta)$ is legal, then its destination
  $(s, t+\delta)$ should also be contained within the
  safety-invariant.
\item[Discrete jumps.] For every state $(s, t)$ that is within the
  safety-invariant, if an $\alpha$-labelled discrete-jump
  $(s,t)\xrightarrow{\alpha}(s',t')$ is legal (i.e., it is allowed due
  to the controller synthesis), then its destination $(s', t')$ should
  also be contained within the safety-invariant.
\item[Guaranteed time progress.] If a mode is bound by an invariant,
  there exists a discrete jump that works on the boundary to enter the
  next mode.
\end{description}

Using the above conditions, readers can observe that we again create
an exists-forall problem for the control of timed systems with
universal variables $s, s'\in \{s_0, \ldots, s_n\}$ and $t,t',\delta
\in \mathbb{R}$. Existential variables are templates and possible
control choices (e.g., restrictions on certain guards or restrictions
on mode invariants). Time progress corresponds to linear arithmetic,
and each mode $s_i$, where $i\in \{0, \ldots, n\}$, is encoded as a
finite bitvector number. Therefore, the whole problem is handled in
\textsf{EFSMT} with a combination of Boolean formula and linear
arithmetic. Notice that here the definition of real-time control
system is slightly more general than timed
automata~\cite{alur:1994:tta}, as the following (somewhat artificial) example
shows.

\paragraph{Example.} Consider a simplified temperature control system in Fig.~\ref{fig:temperature.control}. The system has two modes and has $\alpha, \beta, \gamma, \eta$ as design parameters. The system has a clock $t$ initially set to~$0$. The dynamics of mode~$0$ is described as a simple differential equation $\dot{h}(t)=2$. To find appropriate parameters that satisfies the safety specification, following the template-based approach, we outline the following variables when translating the problem into \textsf{EFSMT}.

\begin{enumerate}
    \item Declare existential variables $\alpha, \beta,  \eta \in \mathbb{R}_{\geq 0}, \gamma\in \mathbb{R}$.
    \item For templates, for mode~0 declare
        $l_{m_0}, u_{m_0}\in \mathbb{R}$ (for lowerbound and upperbound on $h(t)$). Similarly declare $l_{m_1}$, $u_{m_1} \in \mathbb{R}$ for mode~1. Also declare the corresponding primed version.
    \item Use the following universal variables $\textsf{mode}, \textsf{mode}'\in \mathbb{B}$ (for modes; $\textsf{mode}= \textsf{false}$ means that the current location is at mode~0) and  $h, h'\in \mathbb{R}, t, \delta\in \mathbb{R}_{\geq 0}$ (for the change of dynamics and the progress of time).%
\end{enumerate}
Altogether we obtain the following constraints.
\begin{enumerate}
\item The primed version and unprimed version of the invariant should be the same.
\item The initial state is included in the invariant. Introduce the following constraint: $l_{m_0}\leq 100 \leq u_{m_0}$.
\item No risk state is within the safety-invariant. Introduce the following constraints: $80\leq l_{m_0}\leq u_{m_0} \leq 120$ and  $80\leq l_{m_1}\leq u_{m_1} \leq 120$.
\item (Time jump) E.g., the following constraint shows the effect of time jump in mode~1.
      \begin{multline*}
        (\textsf{mode}=\textsf{true}\, \wedge\, \textsf{mode}'=\textsf{true} \\
        \wedge  t \leq 6 \wedge t+\delta\leq 6 \wedge l_{m_1} \leq h \leq u_{m_1}) \rightarrow 
        (l_{m_1} \leq h + \gamma \delta \leq u_{m_1})
      \end{multline*}
      The first two lines specify the assumption that it is a time progress (the evolving of time stays within the invariant), and the third line specifies the guarantee that the effect of time jump is still within the invariant.  While time progresses, $h$
  increases by $\gamma \delta$. As $\gamma \delta$ constitutes a
  nonlinear term in the constraint, a pure linear arithmetic solver is
  unable to handle the problem.

\item (Discrete jump) E.g., the following constraint shows the effect of discrete jump from mode~1 to mode~2.
      \begin{multline*}
        (\textsf{mode}=\textsf{false}\, \wedge\, \textsf{mode}'=\textsf{true} \\
        \wedge  (t \leq \alpha) \wedge  (\beta \leq t \leq 10)  \wedge l_{m_0} \leq h \leq u_{m_0}) \rightarrow 
        (l_{m_1} \leq h \leq u_{m_1})
      \end{multline*}
      The first line specifies the mode change. In the second line, $(t \leq \alpha) \wedge  (\beta \leq t \leq 10)$ specifies the condition for triggering the discrete jump.  $l_{m_0} \leq h \leq u_{m_0}$  and $l_{m_1} \leq h \leq u_{m_1}$ specify the need of staying within the invariant before and after the discrete jump.
\item (Guaranteed time progress) For the first mode to progress, introduce constraint $\beta\leq \alpha\leq 10$. For the second mode to progress, introduce constraint $\eta\leq 6\leq 10$.
\end{enumerate}
Although the
generated constraint is nonlinear, the problem can be solved by
problem discharging. This is because the constraint has one nonlinear
term $\gamma \delta$, where $\gamma$ is an existential variable and
$\delta$ is a universal variable. Using constraint discharging,
\textsf{EFSMT} produces $(\alpha, \beta, \eta, \gamma) = (10, 10,
\frac{-20}{6}, 6)$. Therefore, the synthesized result makes the
temperature control system deterministic: Start from mode~0, continue
heating with ratio~$2$ for 10 seconds and then switch to mode~1. At
mode~1, continue cooling with ratio $\frac{-20}{6}$ for~6 seconds then
switch back to mode~0.

\subsection{BIBO-stability Synthesis and Routh-Horwitz Criterion}\label{sub.sec.routh.horwitz}

\begin{wrapfigure}{r}{0.45\textwidth}
    \centering
     \includegraphics[width=0.4\columnwidth]{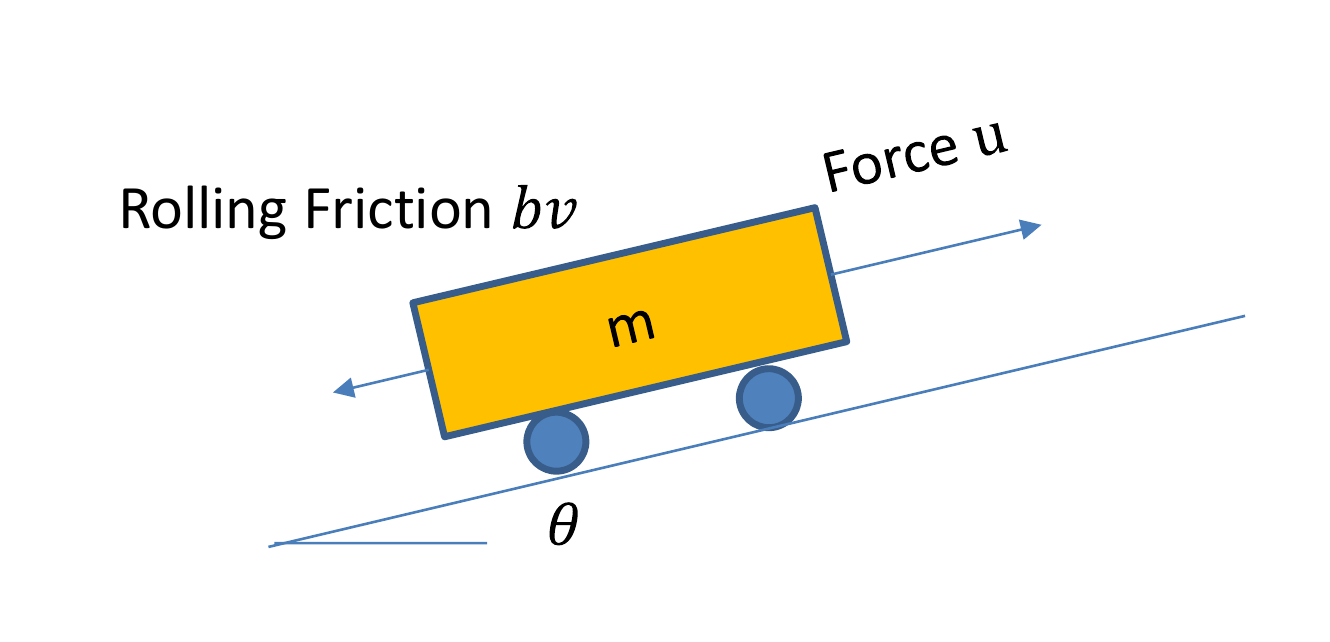}
\vspace{-3mm}
      \caption{The dynamics of a simplified cruise control system.}
     \label{fig:efsmt.sample.cruise.control}
\end{wrapfigure}

\paragraph{Problem description.} Consider a simplified cruise control
system shown in Fig.~\ref{fig:efsmt.sample.cruise.control}.  Given a
constant reference speed $v_r$,
the engine tries to maintain the speed of the vehicle to $v_r$ by applying an
appropriate force $u$\@.  However, in autonomous driving mode, changes in
the slope $\theta$ of the road influence the actual vehicle speed
$v$\@. The rolling friction is proportional to the actual speed with a
constant coefficient $b$\@.

Assume the control of the force is implemented by a
Proportional-Integral (PI) controller with two constants $k_p, k_i$,
i.e., $u = k_p(v_r-v) + k_i \int_{0}^{t} (v_r-v(\tau))d\tau$. Also let
$\theta$ always have a small value ($-10^{\circ} \leq\theta\leq
10^{\circ}$), so we use $\theta$ in replace of $\sin\theta$.  Let $g$
be the gravity constant and $\dot{v}$ be the first derivative of
velocity. If the mass of the vehicle is $m$, we have the following
equation to describe the system dynamics:

\vspace{-5mm}
\begin{multline} \label{eq:dynamics.physics}
m \dot{v} = u - mg\sin \theta - bv 
\cong [k_p(v_r-v) + k_i \int_{0}^{t} (v_r-v(\tau))d\tau] -mg\theta -bv
\end{multline}

We rewrite the equations by setting $v$ to $v_r + \delta$, where
$\delta$ represents the difference between the actual speed and the
reference speed. As $v_r$ does not change over time,
Eq.~\ref{eq:dynamics.physics} is rewritten as:

\begin{equation} \label{eq:dynamics.usage}
m \dot{\delta} = - k_p \delta - k_i \int_{0}^{t} \delta(\tau)d\tau -mg\theta -b(v_r + \delta)
\end{equation}

We define the angle of the road $\theta$ to be the input signal, the velocity difference $\delta$ to be the output signal, and the rest to be internal signals. The Bounded-input-bounded-output (BIBO) stability of the system refers to the requirement that for a bounded angle of the slope, the velocity error compared to the reference $v_r$ should as well be bounded. To ensure BIBO stability of the system, a designer selects appropriate values for control parameters $k_p$ and $k_i$. However, the problem is more complicated when the mass of the vehicle is not a fixed system parameter, but rather a parameter that is within a certain bound to reflect the scenario that 1 to 4 passengers of different weights can be seated in the vehicle during operation. Therefore, the task is to find the set of parameters that ensures BIBO stability for all possible values of the mass. It is important to note that the problem is essentially a game-theoretic setting, as the uncontrollability is reflected at runtime by the variation of passenger loads.

\paragraph{Laplace transform and constraint generation.} For the cruise control problem, we apply a \emph{Laplace transform}\footnote{The Laplace transform of a function $f(t)$, defined for all real numbers $t\geq0$, is the function $F(s):= \mathcal{L}\{f(t)\}(s)= \int_0^{\infty} e^{-st}f(t)dt$.} to create the model to the frequency domain. For simplicity, we neglect friction and  set $b$ to $0$\@. Then the following formula is the corresponding expression of Eq.~\ref{eq:dynamics.usage} in the transferred frequency domain.

\vspace{-2mm}
\begin{equation*} \label{eq:dynamics.transfer.function}
m s \Delta(s) = - k_p \Delta(s) - k_i \frac{\Delta(s)}{s} - mg\Theta(s) 
\end{equation*}

By rearranging the items in the equation, the transfer function of the system is the following form: $\frac{\Delta(s)}{\Theta(s)} = \frac{-mgs}{m s^{2}+k_p s+ k_i}$. Let the denominator of the transfer function of a continuous-time causal system be $Den(s)$, the set of all controllable constants be $C_{ctrl}$, and the set of all uncontrollable constants be $C_{env}$.
Borrowing established results in control theory, BIBO stability is
ensured if all roots of the denominator polynomial have negative real
parts.

\vspace{-2mm}
\begin{multline*}
\exists c_{1}, \ldots, c_{m}\in C_{ctrl} \;\forall e_{1}, \ldots,e_{n}\in C_{env}, \alpha, \beta\in \mathbb{R}: 
Den(\alpha+\beta i)=0 \rightarrow (\alpha<0)
\end{multline*}

Notice that $Den(\alpha+\beta i)=0$ can be rewritten as a conjunction of two constraints where one constraint covers the real part and the other covers the imaginary part. For the cruise control problem, its corresponding algebraic problem can be formulated as the following: $\exists k_p, k_i\; \forall m, \alpha, \beta: m(\alpha+\beta i)^2+k_p(\alpha+\beta i) + k_i = 0 \rightarrow (\alpha < 0)$. By splitting the real part and the imaginary part, we derive the following formula.

\vspace{-2mm}
\begin{multline*}
\exists k_p, k_i\; \forall m, \alpha, \beta: 
(m(\alpha^2-\beta^2)+k_p\alpha + k_i = 0 \, \wedge\,2m\alpha\beta+k_p\beta = 0) 
\rightarrow (\alpha < 0)
\end{multline*}

\paragraph{Routh-Hurwitz criterion.}
The formulation above does not yield bounds on $\alpha$ and $\beta$\@. The \emph{Routh-Hurwitz criterion}~\cite{kuo2003automatic} from the control domain gives sufficient and necessary conditions for stability to hold in a continuous-time system based on analyzing coefficients of a polynomial $\sum^{n}_{k=1} a_k s^k$ without considering $\alpha$ and $\beta$. E.g., if the denominator polynomial is $f(s)=a_4s^4+a_3 s^3+a_2s^2+a_1 s+a_0$, then for all roots to have negative real parts, all coefficients must be greater than $0$, $a_3a_2 > a_4a_1$, and $a_3a_2a_1 > a_4a_1^2 + a_3^2a_0$\@.
The Routh-Hurwitz criterion can be exploited to make \textsf{EFSMT} more efficient.

For the cruise control problem, we have the polynomial $ms^2+k_p s +
k_i$ of second degree. Let $m\in[600, 1200]$, and $k_p, k_i\in [-100,
100]$.  We derive the following simple constraint by applying the
Routh-Hurwitz criterion:

\begin{multline}
\exists k_p, k_i\in [-100, 100]\, \forall m\in [600, 1200]: 
m>0\,\wedge\,k_p>0\,\wedge\,k_i>0
\end{multline}

Therefore, \textsf{EFSMT} returns $k_p, k_i$ by ensuring that they are greater than 0. Often the problem under analysis is described by polynomials of fifth or sixth degree where \textsf{EFSMT} is very useful.

\subsection{Certificate Generation for Lyapunov Functions}

In BIBO stability analysis, the problem is restricted to linear time-invariant (LTI) systems
and the analysis is performed in the frequency domain. \emph{Lyapunov analysis} targets asymptotic stability of nonlinear systems with analysis on the time domain.

\paragraph{Problem description.} Consider the following scalar nonlinear system\footnote{This example is taken from Ex. 4.9 in the book by Astrom and Murray~\cite{astrom2008feedback}.}:
\begin{equation}
    \frac{dx}{dt} = \frac{2}{1+x} - x
\end{equation}

An \emph{equilibrium point} is the point that makes $\frac{dx}{dt} = 0$.\footnote{When $x$ refers to spatial displacement, $\frac{dx}{dt}$
  is the velocity of a moving object and equilibrium point is reached
  when velocity is 0.} The above system has an equilibrium point $x=1$, as $\frac{dx}{dt} = \frac{2}{1+1} - 1=0$. We are interested in certifying the \emph{asymptotic stability} of an equilibrium point, i.e., under small disturbances, whether it is possible to move back to the equilibrium point. For example, for an inverted pendulum, the upright position is \emph{unstable}, as any small disturbance makes the inverted pendulum drop. However, a normal pendulum is stable at its lowest position, as the energy dissipation due to air-friction eventually brings the pendulum back to the low-hanging position.

\paragraph{Lyapunov stability analysis.}  To prove stability, we apply Lyapunov analysis, which targets to find an \emph{energy-like function} $V$ and a \emph{radius} $r$. It then proves that for all points within the bounding sphere whose center is the equilibrium point and the radius is $r$ (except the center where $V(x) = 0$), $V(x) > 0$ and $\dot{V}(x)\leq 0$. Intuitively, as $\dot{V}(x)\leq 0$, the energy dispersion ensures that all points within the sphere stay close to the equilibrium point.

For this problem, we first perform the change of axis by setting $z = x-1$. This sets the equilibrium point to $z=0$.
\begin{equation*}
    \frac{dz}{dt} = \frac{2}{2+z} - z-1
\end{equation*}

The second step is to describe the energy function as templates. Here we use the use $V(z) = az^2$, where $a$ is a constant to be synthesized by \textsf{EFSMT}. Then $\dot{V}(z) = \dot{z}z=2az(\frac{2}{2+z} - z-1) = \frac{-2az(z^2+3z)}{(2+z)}$. Assume our interest is within $-5\leq
z\leq5$. We can then reduce the problem of \emph{Lyapunov stability} to the following:

\vspace{-2mm}
\begin{multline*}
    \exists a, r \,\forall z\in[-5, 5]: (r>0) \wedge ((0<|z|<r)
     \rightarrow (V(z)>0 \wedge \dot{V}(z)\leq0))
\end{multline*}

To process the constraint in \textsf{EFSMT}, we observe that  $\dot{V}(z)$ involves division. $\dot{V}(z)\leq 0$ is equal to the constraint $2az(z^2+3z)(2+z)\geq 0$. For $V(z)>0$, the condition is to have $a>0$. We then derive the following constraint.

\vspace{-2mm}
\begin{multline}\label{eq.solvable.bernstein}
    \exists a>0, r>0 \,\forall z \in [-5, 5]: (0<z<r \, \vee \,-r<z<0 ) \\
    \rightarrow 2az(z^2+3z)(2+z)= 2az^2(z+2)(z+3)\geq 0
\end{multline}

\paragraph{Constraint strengthening.}
Here, we demonstrate the use of constraint strengthening using bitvector theories.
\begin{itemize}
    \item As we know that when $z=0$, the condition  ($\dot{V}(0)=0$) holds, for simplicity we change $(0<z<r \, \vee \,-r<z<0 )$ to $-r <z< r$. After strengthening each conjunction, we derive $z+1_{bit}>-r \wedge z-1_{bit} <r$.

    \item If $2a\geq0$ then $z^2(z+3)(2+z) \geq0$. We have $z\geq0$ or $-2\leq z \leq0$. Strengthening creates $(z-1_{bit}\geq0)$ or $(z-1_{bit}\geq-2 \wedge z+1_{bit}\leq0)$.
    \item If $2a<0$ then $z^2(z+3)(2+z) \leq0$. We have  $z \leq -3$. Strengthening creates $z+1_{bit}\leq -3$.
\end{itemize}

\vspace{1mm}

The constraint in Eq.~\ref{eq.strengthening.Lyanupov.example} is the strengthened condition for using \textsf{EFSMT} with bitvector theories. When setting $1_{bit}$ to be $\frac{1}{32}$, \textsf{EFSMT} returns \textsf{true} with $a = \frac{1024}{32}=32$ and $r = \frac{32}{32}=1$.
Therefore, by using the energy function $V(z)=32z^2$, with bitvector
theories we show that Lyapunov stability is achieved at $x=1$ in a
sphere of radius $1$.

\vspace{-2mm}
\begin{multline}\label{eq.strengthening.Lyanupov.example}
    \exists a, r \in [0, 10]_{BV} \,\forall z \in [-10, 10]_{BV}: 
    (z+1_{bit}>-r \wedge z-1_{bit} <r ) \rightarrow \\
   ((2a\geq 0 \rightarrow ((z-1_{bit} \geq 0)\vee (z-1_{bit}\geq -2 \wedge z+1_{bit}\leq 0)))
																	\wedge
 										(2a<0 \rightarrow (z+1_{bit}\leq -3)))
\end{multline}

\paragraph{Effect of constraint strengthening.} Notice that
Lyapnuov stability guarantees that the system remains near the
equilibrium point, while asymptotic stability guarantees the
convergence toward that point. In this example, due to constraint
strengthening, \textsf{EFSMT} can only prove Lyapunov stability
($V(z)>0 \wedge \dot{V}(z)\leq0$) but not asymptotic stability
($V(z)>0 \wedge \dot{V}(z)<0$), although it also holds for $a=32,
r=1$.

\paragraph{Using JBernstein.}
In Eq.~\ref{eq.solvable.bernstein}, when we follow the first step in strengthening to the change $(0<z<r \, \vee \,-r<z<0 )$ to $-r <z< r$, the newly generated formula already satisfies the shape in Eq.~\ref{eq.forall.assume.guarantee}, thereby is solvable with
\textsf{JBernstein} (as \textsf{F-solver}) and linear arithmetic (for \textsf{E-solver}). With \textsf{JBernstein}, \textsf{EFSMT} returns \textsf{true} with the same radius $r = 1$ but another energy function $V(z)=8z^2$.


\vspace{-2mm}
\section{Using \textsf{EFSMT}}\label{sec.evaluation}

The current implementation of \textsf{EFSMT} uses Yices2 SMT. In addition, we have extended \textsf{JBernstein} with the
following features:
\begin{inparaenum}[(a)]
\item accept constraints with parameterized coefficients (e.g.
  $3x_1+2x_2$),
\item programmatically provide an array of assignments (e.g. $(x_1,
  x_2)=(1,1)$),
\item solve constraints where every coefficient is concretized, and
\item programmatically report the results of validity checking.
\end{inparaenum}
This makes the using of \textsf{JBernstein} into \textsf{EFSMT}
possible. Similar to Yices2, \textsf{EFSMT} offers a C API that
facilitates users to access basic functionalities and to create their
own textual input formats. Fig.~\ref{fig.program} demonstrates the
usage of the API for the simple constraint
\begin{equation*}\label{eq.formula.sample}
\exists x\, \forall y: (0<y<10)\rightarrow (y-2x<7)
\end{equation*}

\lstdefinestyle{customc}{
  belowcaptionskip=1\baselineskip,
  breaklines=true,
  xleftmargin=\parindent,
  language=C,
  showstringspaces=false,
  basicstyle=\scriptsize\ttfamily,
  keywordstyle=\bfseries\color{black},
  commentstyle=\itshape\color{magenta},
  identifierstyle=\color{blue},
  stringstyle=\color{red},
}

\lstset{escapechar=@,style=customc
  ,numbers=left, numberstyle=\tiny, stepnumber=1, numbersep=5pt}
\begin{figure}[t]
\centering
\begin{lstlisting}
#include<iostream>
#include<string>
#include<vector>
#include"efsmt.h"
...
void testExecuteSolver1_LA_LA() {

vector<Variable> existentialVariables;
vector<Variable> universalVariables;
vector<expression> assum;
vector<expression> guar;
vector<expression> cond;

// Existential variables
Variable x = { "x", EFSMT_VAR_REAL };
existentialVariables.push_back(x);

// Universal variables
Variable y = { "y", EFSMT_VAR_REAL };
universalVariables.push_back(y);

// Assumption
assum.push(insertAssignment("y>0","GT","y","0"));
assum.push(insertAssertion("y>0"));
assum.push(insertAssignment("y<10","LT","y","10"));
assum.push(insertAssertion("y<10"));

// Guarantees
guar.push(insertAssignment("tmp","MUL","2","x"));
guar.push(insertAssignment("tmp2","SUB","y","tmp"));
guar.push(insertAssignment("y-2x<7","LT","tmp2","7"));
guar.push(insertAssertion("y-2x<7"));

// Conditions
EFSMTProblem prob;
prob.existentialVariables = existentialVariables;
prob.universalVariables = universalVariables;
prob.assumptions = assum;
prob.guarantees = guar;
prob.conditions = cond;
prob.problemType = EFSMT_PROB_LA_LA;
prob.solverOption = EFSMT_FULL;
executeEFSMTSolver(prob, EFSMT_SOLVER_YICES2,false);
}
\end{lstlisting}
\label{fig.program}
\caption{Encode constraint $\exists x\, \forall y: (0<y<10)\rightarrow
  (y-2x<7)$ using the API of \textsf{EFSMT}\@.}
\end{figure}

\begin{wrapfigure}{r}{0.55\textwidth}
\includegraphics[width=0.25\columnwidth]{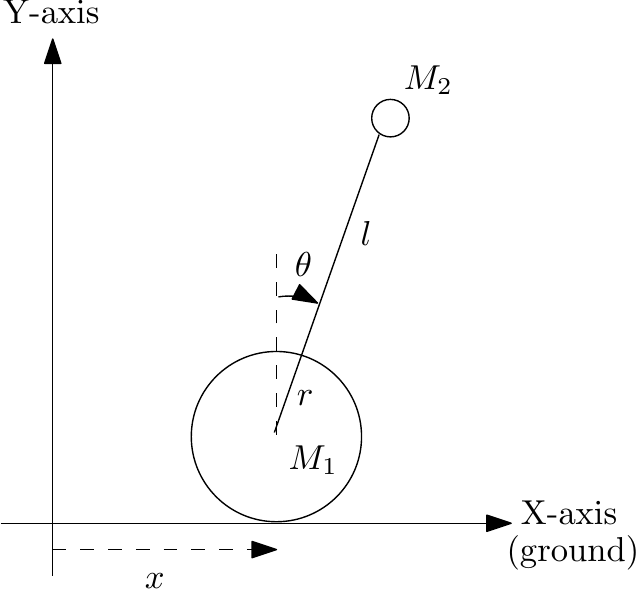}
  \caption{A physical model of a wheeled inverted pendulum .}
 \label{fig:WIP}
\end{wrapfigure}

 First, we declare two vectors to store existential and universal
  variables. Line~15 declares variable $x$ in the domain of
  reals. Line~16 categorizes $x$ as an existential variable.
Then we define three vectors \texttt{assum} (stores assumptions
  in universal variables), \texttt{cond} (stores conditions in
  existential variables), and \texttt{guar} (the general
  constraint). Constraints stored in each vector are conjuncted.
  Therefore, $(\texttt{assum} \rightarrow \texttt{guar})
  \wedge \texttt{cond}$ forms the specified quantifier-free
  constraint. Admittedly, all constraints can be described in the
  \texttt{guar}-vector. The separation is for performance
  considerations: For example, separating \texttt{cond} from the
  general constraint allows the \textsf{F-solver} to omit constraints
  specified in \texttt{cond}.
 There are two types of actions: \texttt{insertAssignment()} that
  creates intermediate terms and \texttt{insertAssertion()} that
  specifies a term which evaluates to either \textsf{true} or
  \textsf{false}. At line~29, $2 \times x$ is created and stored in
  variable \texttt{tmp} and line~32 creates the "\texttt{y-2x<7}"
  constraint added to the \texttt{guar} vector.
Line~41 specifies the problem solving type as
  \texttt{EFSMT\_PROB\_LA\_LA}, meaning that both \textsf{E-solver}
  and \textsf{F-solver} are handled by linear arithmetic. Line~42
  enforces complete search. Line~43 invokes the solver with Yices2 as
  the underlying engine.
The full API specification and documentation is included in the \texttt{efsmt.h} header file.




\paragraph{Performance.} We briefly summarize preliminary results concerning the
performance of \textsf{EFSMT}. For nonlinear constraint checking, due to advantages
offered by \textsf{JBernstein}, the performance is considerably
fast. For example, in the PVS test suite, \textsf{JBernstein} solves
problems (in the best case) two to three orders of magnitude faster
than existing tools such as QEPCAD~\cite{Brown:2003:QBP:968708.968710}
and REDLOG~\cite{Dolzmann:1997:RCA:261320.261324}. Short query
response time makes the counter-example guided approach
applicable. For problems with only Boolean variables (2QBF), the
introduction of multiple instances (e.g., set $\alpha=2$ or $3$)
ameliorates performance nearly linear to the used number (when number is small).
Because we do not modify the underlying code structure of Yices2, we are
unable to integrate known tricks that are used in 2QBF solving,
such as Plaisted-Greenbaum transformation~\cite{PG86}. We have
independently implemented another 2QBF solver using SAT4J~\cite{SAT4J}
that utilizes partial assignment and contexts. We compare it with the
QBF solver QuBE++~\cite{giunchiglia2004qube++} by disabling its
preprocessing ability (i.e., to perform simplification and generate
formulas with fewer variables) to compare the performance on the core
engine. Not surprisingly, as our implementation extends the work
in~\cite{janota2011abstraction} which has demonstrated its superiority
over QuBE++, the solver is faster even without our optimization.

\paragraph{Case study: wheeled inverted pendulum.} We outline a
concrete example in modeling and parameter synthesis that ensures
stability of a wheeled inverted pendulum - a two-wheeled
Segway\footnote{http://www.segway.com/} implemented with Lego
Mindstorm\footnote{http://mindstorms.lego.com/} and
RobotC\footnote{http://www.robotc.net/}.

\paragraph{Design and Assumptions.} During the design process, two wheels
  are locked to allow only forward and backward movement. We assume
  that wheels are always in contact with the ground and experience
  rolling with no slip. Furthermore, we consider no electrical and
  mechanical loss. Finally, the inverted pendulum is equipped with a
  Gyro-meter to measure angular displacement and velocity.

\paragraph{Open system.}
The graphical model of the open system with the
  above assumptions is shown in Fig.~\ref{fig:WIP}. By using the
  Lagrange method (generalized Newton dynamics), we derive the
  dynamics of the system to the following equation\footnote{Due to
    space limits, the complete derivation is omitted.}:
  \begin{equation*}
    M_2 l \ddot{x} \cos \theta + (J_2 +M_2 l^2)\ddot{\theta} -M_2 gl \sin\theta = 0
  \end{equation*}
  where $J_1$ and $J_2$ are rotation inertia for the wheel and
  object (represented as the upper circle), $g$ is the Newtonian gravity
  constant, $\ddot{x}$ is the second derivative of displacement, and
  $M_1$, $M_2$ are masses of the wheel and the object.

\paragraph{Control.}
Let $\tau$ be the provided torque, the term we want to
  control to avoid falling. Then the rotational acceleration generated
  by the torque on the wheel is given by $\ddot{x} = r \alpha$, where
  $\alpha = \frac{\tau}{J_1}$. Assume that the inverted pendulum is
  initially placed vertically ($\theta = 0$)  and will experience only very small disturbance.
  We apply small-angle approximation to let $\sin \theta
  \approx \theta$ and $\cos \theta \approx 1$. After simplification,
  the following equation is generated:
  \begin{equation}
    M_2 l r \frac{\tau}{J_1} + (J_2 +M_2 l^2)\ddot{\theta} -M_2 gl \theta = 0
  \end{equation}
  From this equation we observe that when no torque is applied
  ($\tau=0$), state
    $(\theta, \dot{\theta}) = (0,0)$ is an equilibrium
    point for the inverted pendulum. However, any small displacement (i.e., $\theta\neq 0$)
  creates $\ddot{\theta}$ and makes the pendulum fall. Let the
  controller be implemented with Proportional (P) or
  Proportional-Derivative (PD) controllers. Then we have two unknown
  parameters $k_p$, $k_i$ for synthesis. Implementing a PD controller
  replaces $\tau$ by $k_p(\theta - 0)+k_d \dot{\theta}$.

\paragraph{Synthesis.}
 In this problem we need to synthesize control
  parameters $k_p$, $k_d$ which are represented as existential
  variables. To prove Laynupov stability, we need to find parameters
  for the energy function. Let $x_1 = \theta$ and $x_2 =
  \dot{\theta}$. For energy functions, we use templates such as $V=a
  (x_1)^2 + b (x_2)^2$.   We use existential variables $\overline{x_1}, \overline{x_2}$
  to form the bounding sphere. States $(x_1, x_2)$ and
  environment parameters that need to be tolerated are expressed by
  universal variables. We also let $M_2$, $l$ be universal variables
  that range within bounded intervals to capture modeling imprecision;
  we assume that the fluctuation of $M_2$ and $l$ are sufficient for
  capturing the real dynamics of the pendulum shown on the left of
  Fig.~\ref{fig:WIP}. Finally, we pass the following constraint to
  \textsf{EFSMT} to synthesize control parameters for asymptotic
  stability.
  \begin{multline*}
    \exists \overline{x_1}, \overline{x_2}>0 \,\exists k_p, k_d \,\exists a, b>0 \,\forall x_1, x_2, M_2, l :\\
    (|x_1|< \overline{x_1} \,\wedge\, |x_2|< \overline{x_2})
    \rightarrow (V(x_1, x_2) > 0\, \wedge \,\dot{V}(x_1, x_2) < 0)
  \end{multline*}
  In both P and PD controllers, \textsf{EFSMT} fails to find a witness
  for asymptotic stability $\dot{V}(x_1, x_2) < 0$ with the initial
  condition $(\theta, \dot{\theta})=(k, 0)$ where $k\neq 0$, as these
  states makes $\dot{V}(x_1, x_2) = 0$.
  Therefore, with Lyapunov theorem, the solver can at best prove
  Lyanupov stability (i.e., $\dot{V}(x_1, x_2) \leq 0$). Stronger
  results, such as Krasovski-Lasalle
  principle~\cite{astrom2008feedback}, are needed to derive asymptotic
  stability when $\dot{V}$ is negative semidefinite.

\vspace{-2mm}

\section{Related Work}\label{sec.related}

\textsf{EFSMT} differs from pure theory solvers such as Cylindrical Algebraic Decomposition
(CAD)~\cite{collins1991partial} tools (e.g.,
QEPCAD~\cite{Brown:2003:QBP:968708.968710} and
REDLOG~\cite{Dolzmann:1997:RCA:261320.261324}) or Quantified Boolean
Formula (QBF) solvers (e.g., QuBE++~\cite{giunchiglia2004qube++}) in
that it allows combination of subformulas with variables in different
domains into a single constraint problem. From a design perspective,
this allows to model control and data manipulation simultaneously (for
example, \textsf{EFSMT} can naturally encode the hybrid control problem stated
in Section.~\ref{sub.sec.timed.systems}). Compared to SMT solvers such
as Yices2, Z3~\cite{de2008z3}, or
openSMT~\cite{bruttomesso2010opensmt}, \textsf{EFSMT} can analyze more
expressive formulas with one quantifier alternation. Furthermore, the
nonlinear arithmetic is based on Bernstein polynomials~\cite{MN12} and
other assisting techniques (discharging, strengthening) rather than
CAD. Admittedly, CAD can solve problems with arbitrary quantifier
alternation. However, problems under investigation in \textsf{EFSMT}
are more restricted, as we only have one quantifier
alternation. Overall, our methodology applies verification techniques
(such as CEGAR, abstract interpretation) to guide the process. Our
counter-example based approach can be viewed as a technique of
CEGAR. The use of two solvers in our solver is borrowed from the technology in the SAT
community~\cite{2QBFViaSAT}. It is also an extension from recent results in solving QBF via
abstraction-refinement~\cite{janota2011abstraction}. However, our
method is combined with infeasibility test (that uses widening) to
fully utilize the capability of two separate solvers. In addition, our
proposed optimization techniques (e.g., the effect of partial
assignment) are never considered by these works.

For reduction problems presented in this work, safety orchestration of
component-based systems using priorities was first proposed by Cheng
et al.~\cite{cheng:vissbip:2011}. The algorithm is based on a
heuristic that performs bug finding and fixing. We extend the work by
using a template-based reduction that allows to easily encode
architectural constraints and other artifacts in distributed execution
(e.g., knowledge). Safety control for timed systems first appeared in
the work by Maler, Pnueli and Sifakis~\cite{maler1995synthesis}. Tools
such as UPPAAL-Tiga~\cite{behrmann2007uppaaltiga} allow to synthesize
strategies for timed games. \textsf{EFSMT} uses a template-based
approach, meaning that the synthesized safety-invariant is fixed in
its shape. Therefore, the goal is not to find a controller that
maintains maximal behavior. However, as demonstrated in the example
(Section~\ref{sub.sec.timed.systems}), constraint encoding allows
synthesis on a wider application not restricted to timed systems. For
stability control~\cite{astrom2008feedback}, Lyapunov functions are in
general difficult to find. Our reduction to \textsf{EFSMT} searches
for feasible parameters when the shape of the Lyapunov function is
conjectured. It can be used as a complementary technique to known
methods in control domain that systematically searches for candidate
templates (e.g., sum-of-square
methods~\cite{PrajnaSOSTOOLS}). Finally, for the analysis in the
frequency domain (Section~\ref{sub.sec.routh.horwitz}), graphical or
numerical methods (such as Bode diagrams~\cite{kuo2003automatic}) are
often used to decide the position of a root in the complex
plane. These methods are applied when parameters are fixed. Other
graphical methods such as Nyquist plot~\cite{kuo2003automatic} are
used for deciding parameterized behavior. Algebraic or symbolic
methods that extend Routh-Horwitz criterion include the well-known
Kharitonov's theorem~\cite{bhattacharyya1995robust} which allows to
detect stability for polynomials with parameters of bounded range. In
other words, if values of design parameters are provided by the
$\exists$-solver, an implementation of Kharitonov's theorem can also
act as a $\forall$-checker for BIBO stability.
Gulwani and Tiwari verify hybrid systems~\cite{Tiwali2008CAV}
with exists-forall quantified first-order fragment of propositional
combinations over constraints, and the technology is based on a
quantifier elimination procedure. As demonstrated in our example, we use the
technique to do synthesis of hybrid control systems, and our method is based on CEGAR.
In addition, we show a richer set of problems that can be encoded within the framework such as Lyanupov
certificate synthesis. We also show that progress can be ensured
without another quantifier alternation by carefully trimming the synthesized
strategy structure to be priorities.

\vspace{-2mm}

\section{Conclusion}\label{sec.conclusion}

\textsf{EFSMT} extends propositional \textsf{SMT} formulas with one top-level
exists-forall  quantification. We have demonstrated, by means of reducing a variety of design
problems, including the stability of control systems and orchestration of system components,
that \textsf{EFSMT} is an adequate logical framework for the design and the analysis
of cyber-physical systems. The \textsf{EFSMT} fragment of first-order logic
is expressive enough to allow strategy finding for safety games, while strategies for other properties
can be derived by either restricting the structure (e.g., the use of priorities to ensure progress) or game transformation
(e.g., bounded synthesis~\cite{boundedSynthesis} that transforms LTL synthesis to safety games via behavioral
restrictions). We also propose an optimized verification procedure for solving \textsf{EFSMT} based on
state-of-the-practice \textsf{SMT}  solvers and the use of Bernstein polynomials for  solving
nonlinear arithmetic constraints.  Although we have restricted ourselves to arithmetic constraints,
the approach is general in that rich combinations of theories, as supported by \textsf{SMT} solvers,
can readily be incorporated. Future extensions of the \textsf{EFSMT} proof procedure
are mainly concerned with performance and usability enhancements, but also with completeness.

\end{document}